\documentclass[11pt]{article}
\usepackage{moriond,epsfig}

\def\Journal#1#2#3#4{{#1} {\bf #2}, #3 (#4)}

\def\AAP{\em Astron. Astrophys.}
\def\MN{\em Mon. Not. R. Astr. Soc.}
\def\PASJ{\em Publ. Astr. Soc. Japan.}
\def\APJ{\em Astrophys. J.}
\def\be{\begin{equation}}
\def\ee{\end{equation}}
\def\bea{\begin{eqnarray}}
\def\eea{\end{eqnarray}}

\begin{document}
\vspace*{4cm}
\title{DISTRIBUTION OF BARYONIC AND NON BARYONIC MATTER IN CLUSTERS OF GALAXIES}

\author{A. CASTILLO-MORALES and S. SCHINDLER }

\address{Astrophysics Research Institute, Liverpool John Moores
University\\Twelve Quays House, Birkenhead CH41 1LD, United Kingdom}

\maketitle\abstracts{
We present the analysis of baryonic and non-baryonic matter
distributions in a sample of eleven nearby clusters (0.03 $<$ z $<$
0.09) with temperatures between 4.4 and 9.4 keV. These galaxy clusters have been studied in detail using X-ray
data and global physical properties have been determined.
Correlations between these quantities have been analysed and compared
with the results for distant clusters. We found an interesting dependence between the relative gas extent
(expressed as the ratio of gas mass fractions at $r_{500}$ and
$0.5\times r_{500}$) and the total cluster mass. The extent of the gas
relative to the extent of the dark matter tends to be larger in less
massive clusters. This dependence might give us some hints about
non-gravitational processes in clusters.}

\section{Introduction}

Clusters of galaxies are fascinating objects. They represent important
structures in the universe and the understanding of the physical
processes during their formation is an attractive task. The study
of correlations between different physical properties as X-ray
luminosity, temperature, total mass, gas mass and relative gas extent
enable us to draw conclusions about their formation processes,
about the thermal history of clusters and hence about cosmological parameters.
Some authors have studied cluster samples to search for these
fundamental relations (e.g., Fukazawa 1997~\cite{fukazawa}; Allen $\&$
Fabian 1998~\cite{allenfab1}; Arnaud $\&$ Evrard 1999~\cite{arnaud};
Ettori $\&$ Fabian 1999~\cite{ettofab}; Horner et
al. 1999~\cite{horner}; Jones $\&$ Forman 1999~\cite{jonesforman};
Mohr et al. 1999~\cite{mohr}; Schindler 1999~\cite{sab}; Neumann $\&$ Arnaud 1999~\cite{neumann}). 

We selected a sample of nearby clusters of galaxies in which an accurate
total mass determination is possible, i.e with relaxed and symmetric
morphologies, good temperatures measurements and good surface
brightness profiles. All the masses have been calculated in a
consistent way and within equivalent volumes for all the clusters. We
analyse relations between different properties and compare them with
the relations in the more distant sample (0.3 $<$ z $<$ 1.0) by Schindler 1999~\cite{sab}.
We included in the comparison two more distant clusters: RBS797
(z=0.35) analysed with a Chandra observation by Schindler et al.
2001~\cite{sab2} and the cluster RXJ0849+4452 (z=1.26) where we have calculated
the total and gas mass using the parameters from the Chandra
data analysis by Stanford et al. 2000~\cite{stan}. Our preliminary results are
presented here. The cosmology used is: $H_{0}$ = 50 km/s/Mpc and $q_{0}$ = 0.5.

\section{Data Reduction and Analysis}

X-ray imaging data retrieved from the ROSAT archive is used \footnote{\footnotesize{http://www.xray.mpe.mpg.de/rosat/archive}} to
determine the surface brightness profiles of the clusters. For each
cluster a ROSAT PSPC image was reduced using the standard analysis
with EXSAS software. In order to maximize the signal-to-noise ratio,
we use the hard energy band (0.5-2.0 keV). The images were corrected
for exposure variations and telescope vignetting using exposure maps generated with MIDAS/EXSAS software.

To calculate the gas density profiles the standard $\beta$-model
(Cavaliere $\&$ Fusco-Fermiano 1976~\cite{cav}) has been used.
\begin{equation} 
\rho_{gas}(r)=\rho_{0}\left[1+\frac{r^{2}}{r_{c}^{2}}\right]^{-\frac{3}{2}\beta}
\label{eq:gas}
\end{equation}
We generated radial surface brightness profiles in 
concentric annuli (centred on the emission maximum in the
cluster) excluding obvious point sources manually. The observed
profiles are fitted with a $\beta$-model plus background:
\begin{equation}
S(b)=S_{0}\left[1+\frac {b^{2}}{r_{c}^{2}}\right]^{-3\beta+\frac {1}{2}}+B.
\label{eq:surf}
\end{equation}

As the overall $\beta$-model fit is a poor description of the central
region of some clusters where excess emission is observed (due to a
cooling flow or a central point source) we minimised the
reduced $\chi^{2}$ by excluding the central bins from the fit.
The best fit $\beta$-model was determined by excluding the profile
within the cooling radius (Peres et al. 1998~\cite{peres}, Allen $\&$ Fabian
1997~\cite{allenfab}, White et al. 1997~\cite{white}), taking into
account the errors due to the uncertainty of the cooling radius in the errors of the fit parameters.

With the assumption of hydrostatic equilibrium
and spherical symmetry cluster masses can be derived directly from
X-ray observations through the gas density gradient, the gas
temperature gradient and the gas temperature itself:

\begin{equation}
M(<r)=-\frac{kr}{\mu
m_{p}G}T_{gas}(r)\left (\frac{d\ln \rho_{gas}(r)}{d\ln r}+\frac{d\ln T_{gas}(r)}{d\ln r}\right).
\label{eq:mass}
\end{equation}

Weighted ASCA temperatures for our targets were taken from (Markevitch, et al. 1998~\cite{mark},
White 2000~\cite{white2}). We choose the temperatures obtained when
the central cluster regions are excluded to avoid the complication due
to the additional cool emission component.
In this way we determine the spatial distributions of gas mass and total mass from
X-ray surface brightness and ASCA temperatures assuming isothermality (Irwin $\&$ Bregman 2000~\cite{irwin}). This
provides also estimates for quantities like e.g the baryonic fraction
and the gas extent relative to the dark matter distribution.

\section{Results and discussion}
Having acquired the gravitational mass profiles for the clusters
sample, it is now important to determine the radius within which to
calculate the cluster mass. As the mass of a cluster is increasing
with radius, masses can only be compared when derived within
equivalent volumes. Simulations by Evrard et al. 1996~\cite{evrard} have
shown that the assumption of hydrostatic equilibrium is generally valid 
within a radius $r_{500}$, where the mean gravitational mass density
is equal to 500 times the critical density
$\rho_{c}(z)=3H_{0}^{2}(1+z)^{3}/8 \pi G$. We calculated $M_{tot}$ and
$M_{gas}$ at $r_{500}$. Some of the preliminary results in the analysis of the nearby cluster
sample are presented here.

\subsection{Gas mass fraction}

We find that inside each cluster the gas mass fraction increases
outwards (see figure {\ref{fig:fgasrad}) implying that the gas distribution is more extended than dark
matter (see section \ref{subsect:gasextent} for a detailed discussion). A low $\Omega$ is required to explain the high gas mass fraction of $<f_{gas}>_{r_{500}}=0.16\pm0.02$, when compared to the baryon fraction predicted by primordial
nucleosynthesis. From cluster to cluster we find large variations in
the gas mass fraction as Ettori $\&$ Fabian 1999~\cite{ettofab} found.
The gas mass fraction at $r_{500}$ ranges from
$0.121^{+0.014}_{-0.015}$ for cluster A644 to
$0.20^{+0.04}_{-0.03}$ for cluster A3112. These variations have some
implications on cluster formation because it probably reflects the
distribution of baryonic and non-baryonic matter in the early
universe. If all the clusters had originally the same gas mass
fraction and all the differences came later by different amounts of
gas released by the cluster galaxies, larger metallicities in clusters
with high gas mass fraction would be expected. But this not observed
(Schindler 1999~\cite{sab}). Therefore the difference must be caused at least partially by the primordial distribution or baryonic and non-baryonic matter.

\begin{figure}
\centering
%\rule{5cm}{0.2mm}\hfill\rule{5cm}{0.2mm}
%\vskip 1cm
%\rule{5cm}{0.2mm}\hfill\rule{5cm}{0.2mm}
\psfig{figure=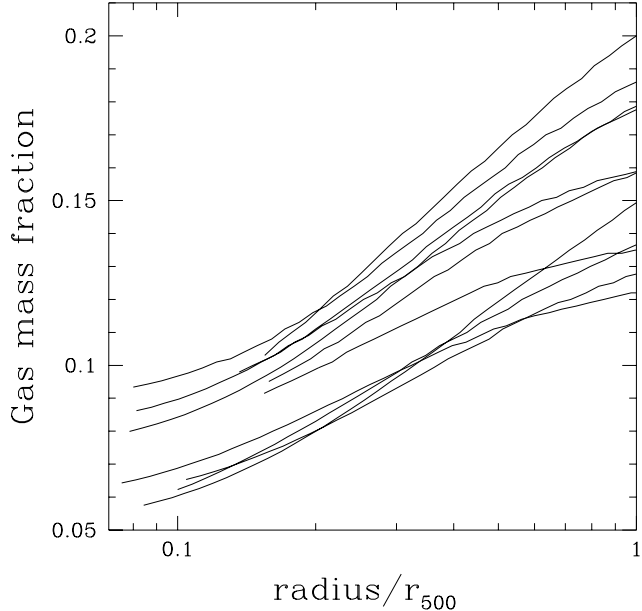,width=10.cm, clip=}
\caption{Gas mass fraction profiles derived for the nearby cluster
sample. Profiles are  plotted from the minimum radius fitted
in the $\beta - $model. 
\label{fig:fgasrad}}
\end{figure}

\begin{figure}
\centering
%\rule{5cm}{0.2mm}\hfill\rule{5cm}{0.2mm}
%\vskip 3cm
%\hskip 5cm
%\rule{5cm}{0.2mm}\hfill\rule{5cm}{0.2mm}
\psfig{figure=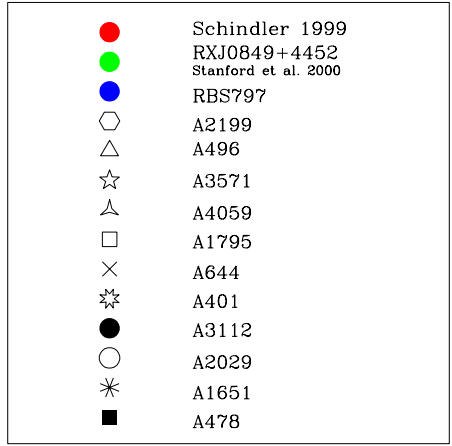,width=7.cm,clip=}
%\special{hoffset=3cm voffset=2cm}
\caption{Symbols used for various clusters in the figures.
\label{fig:symbols}}
\end{figure}

To test whether there is any dependence on redshift with the gas mass
fraction we plot this quantity versus redshift including the results
from Schindler 1999~\cite{sab} for distant clusters in red circles
and the other two distant clusters: RBS797 (blue circle) and
RXJ0849+4452 (green circle) (see figure \ref{fig:fgasz}).
The mean value for the nearby sample is $<f_{gas}>_{r_{500}}=0.16\pm0.02$ being
similar within the errors to the mean value for the distant sample
$<f_{gas}>_{r_{500}}=0.18$. Hence we see no clear trend in the gas mass fraction
with redshift in these data implying an early gas
presence in the cluster (only the most distant cluster
RXJ0849+4452 shows a lower value of 0.11). This contradicts the results by
Ettori$\&$Fabian 1999~\cite{ettofab} where evolution of the gas mass
fraction is found in a nearby sample. Matsumoto et
al. 2000~\cite{matsu} found no clear evidence of evolution for the
clusters at z $<$ 1.0.

\begin{figure}
\centering
%\rule{5cm}{0.2mm}\hfill\rule{5cm}{0.2mm}
%\vskip 1cm
%\rule{5cm}{0.2mm}\hfill\rule{5cm}{0.2mm}
%\hskip 2cm
%\vskip -2cm
\psfig{figure=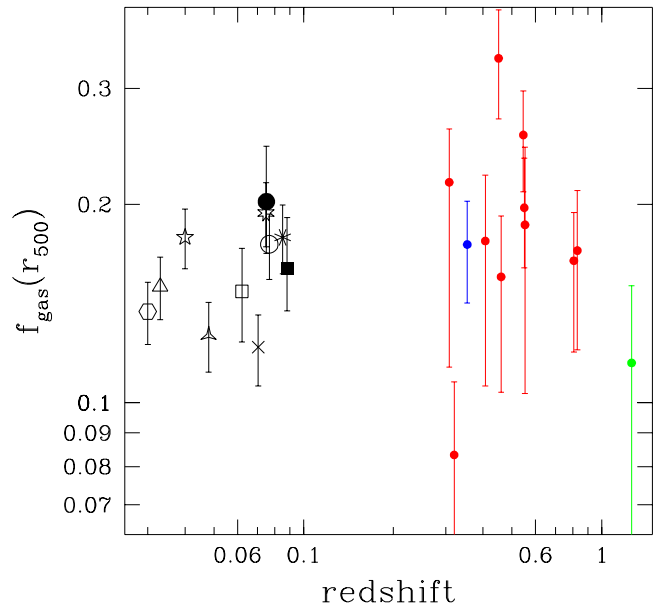,width=10.cm,clip=}
\caption{Gas mass fraction at radius $r_{500}$ versus redshift.
\label{fig:fgasz}}
\end{figure}

\begin{figure}
\centering
%\rule{5cm}{0.2mm}\hfill\rule{5cm}{0.2mm}
%\vskip -8cm
%\hskip 8 cm
%\rule{5cm}{0.2mm}\hfill\rule{5cm}{0.2mm}
\psfig{figure=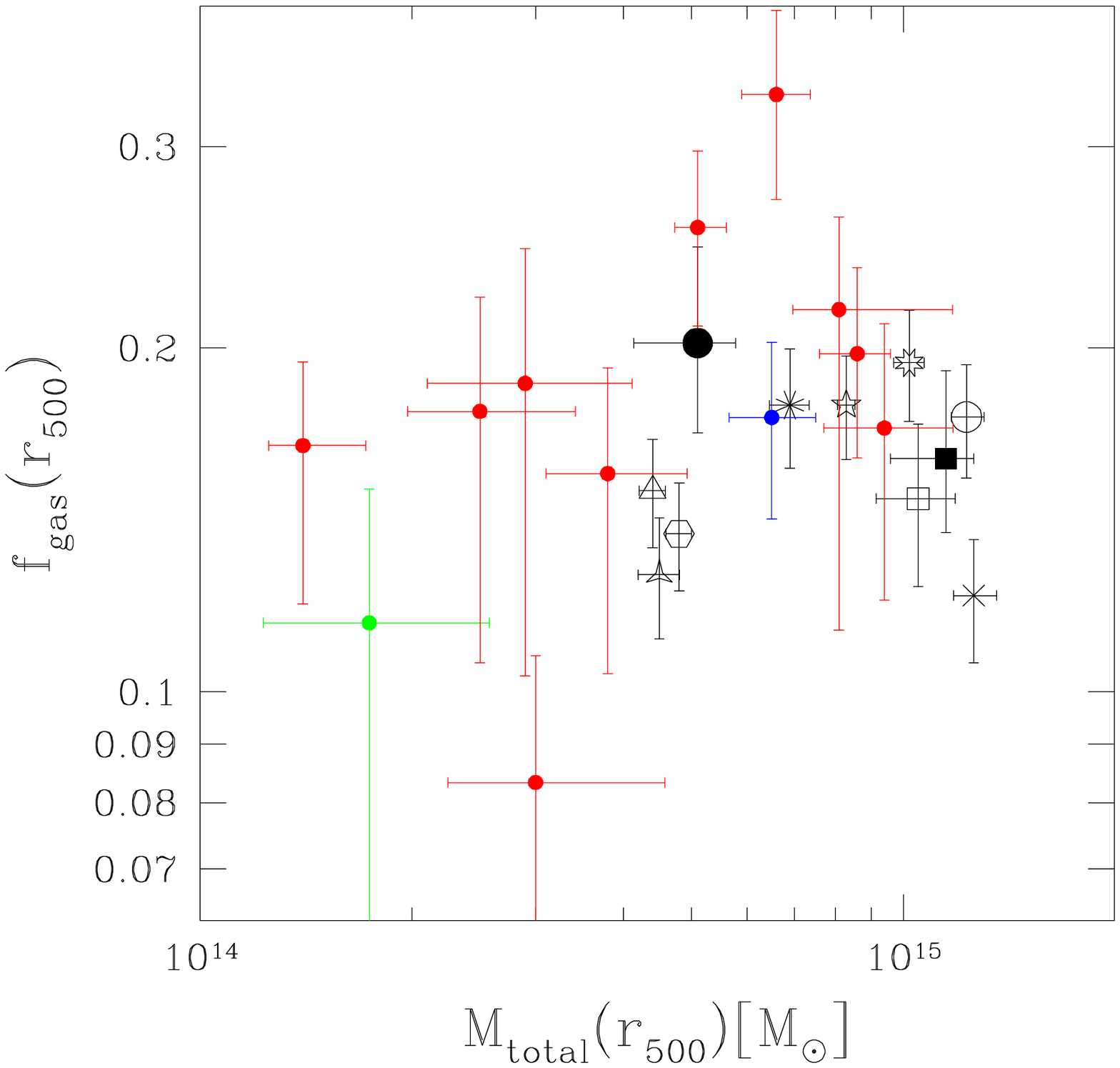,width=10.cm, clip=}
\caption{Gas mass fraction versus total mass at radius $r_{500}$.
\label{fig:fgasMt}}
\end{figure}

We have also looked for trends of $f_{gas}$ with the total cluster mass
$M_{tot}(r_{500})$, see figure \ref{fig:fgasMt}. The gas mass
fraction does not show a clear correlation with the
total mass at radius $r_{500}$. Instead we find the similar trends as
found by Reiprich $\&$ B\"ohringer 1999~\cite{reiprich}: for cluster
masses greater than $5\times 10^{14}M_{\odot}$ the $f_{gas}(r_{500})$
seems to decrease with the total cluster mass and for masses lower than the
above limit $f_{gas}(r_{500})$ tends to increase with total cluster mass.

\subsection{Gas extent}
\label{subsect:gasextent}

As mentioned before the gas mass fraction is not constant with
radius. We compare the gas mass fraction at radius $r_{500}$ with the gas mass
fraction at $0.5\times r_{500}$ in each cluster. The mean gas mass
fraction at $0.5\times r_{500}$ is 0.13, i.e. smaller than the mean of
0.16 at $r_{500}$.
The ratio of these fractions is a measure of how fast the gas mass fraction is increasing with radius and as such a measure for the extent of the gas distribution with respect to the dark mater distribution.

For all the clusters in our sample the relative gas extent E is larger
than 1 (see figure \ref{fig:nearbyE_Mt}). This means that in general the gas
distribution is more extended than the dark matter in agreement with
results from other authors for nearby and distant cluster. The distant
sample by Schindler 1999~\cite{sab} is also shown in figure
\ref{fig:E_Mt} for comparison.
\begin{figure}
\centering
%\rule{5cm}{0.2mm}\hfill\rule{5cm}{0.2mm}
%\vskip 1cm
%\rule{5cm}{0.2mm}\hfill\rule{5cm}{0.2mm}
\psfig{figure=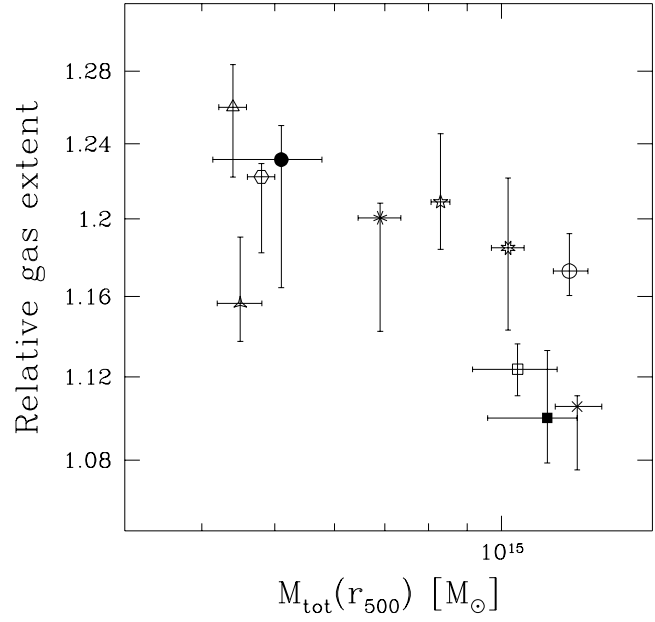,height=10cm, clip=}
\caption{Ratio of gas mass fraction at $r_{500}$ and $r_{500}/2$ as a
measure for the relative gas extent of the gas distribution versus
total cluster mass at $r_{500}$ for the nearby sample.
\label{fig:nearbyE_Mt}}

\end{figure}
\begin{figure}
\centering
%\rule{5cm}{0.2mm}\hfill\rule{5cm}{0.2mm}
%\vskip 1cm
%\rule{5cm}{0.2mm}\hfill\rule{5cm}{0.2mm}
\psfig{figure=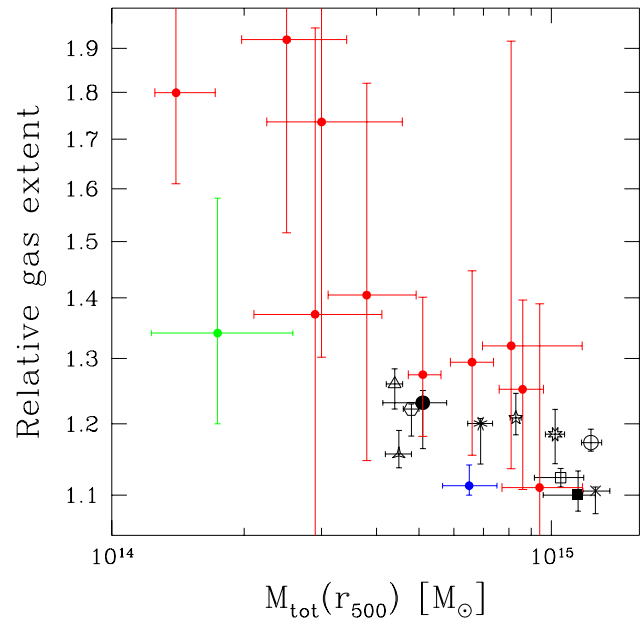,height=10cm, clip=}
\caption{Ratio of gas mass fraction at $r_{500}$ and $r_{500}/2$ as a
measure for the relative gas extent of the gas distribution versus
total cluster mass at $r_{500}$.
\label{fig:E_Mt}}
\end{figure}

In the nearby sample this relative gas extent E shows a mild
dependence on the total mass (see figure \ref{fig:nearbyE_Mt}). Clusters with larger masses tend to have smaller relative gas extents (similar dependence confirmed by Reiprich $\&$ B\"ohringer 1999~\cite{reiprich}).

The distant and low mass cluster RX J0849+4452 seems to follow the same trend
as the nearby sample showing a higher value for the relative gas
extent. The cluster RBS797 lies in the trend as well.
 
This trend can be explained by the physical processes in the gas,
which are assumed to be responsible for the increase of gas mass
fraction with radius like e.g. energy input by supernovae driven
galactic winds. If gas  is placed artificially into a model
cluster potential in hydrostatic equilibrium the distributions of gas
and dark matter have the same slope at radii larger than the core
radius, therefore one would expect a priori a ratio $E\approx1$. It
might be that this additional heat input affects low mass clusters
more that massive clusters, so that a massive cluster can maintain a
ratio E = 1 while in the smaller clusters the gas is becoming more and
more extended.

\section{Further work}

The nearby sample of clusters need to be extended to lower temperature
clusters to confirm the trend of more extended gas distribution in low
mass clusters. At the same time the sample is going to be more complemented by CHANDRA and XMM data.

\section*{Acknowledgments}
We would like to thank the organizers of the meeting for the financial
support. We thank Javier Campania for the help printing the poster.

\end{document}